\documentclass[english,amsmath,amssymb,aps,prl,preprint,superscriptaddress]{revtex4}
\usepackage[latin9]{inputenc}
\usepackage{color}
\usepackage{bm}
\usepackage{amstext}
\usepackage{amssymb}
\usepackage{graphicx}

\makeatletter
\@ifundefined{textcolor}{}
{%
 \definecolor{BLACK}{gray}{0}
 \definecolor{WHITE}{gray}{1}
 \definecolor{RED}{rgb}{1,0,0}
 \definecolor{GREEN}{rgb}{0,1,0}
 \definecolor{BLUE}{rgb}{0,0,1}
 \definecolor{CYAN}{cmyk}{1,0,0,0}
 \definecolor{MAGENTA}{cmyk}{0,1,0,0}
 \definecolor{YELLOW}{cmyk}{0,0,1,0}
}

\makeatother

\usepackage{babel}

\usepackage{caption}
\DeclareCaptionFormat{myformat}{\fontsize{9}{11}\selectfont#1#2#3}
\begin{document}

\title{Trade-off shapes diversity in eco-evolutionary dynamics}

\author{Farnoush Farahpour}

\affiliation{Bioinformatics and Computational Biophysics, University of Duisburg-Essen,
Germany}

\author{Mohammadkarim Saeedghalati}

\affiliation{Bioinformatics and Computational Biophysics, University of Duisburg-Essen,
Germany}

\author{Verena Brauer}

\affiliation{Biofilm Center, University of Duisburg-Essen, Germany}

\author{Daniel Hoffmann}

\affiliation{Bioinformatics and Computational Biophysics, University of Duisburg-Essen,
Germany}

\affiliation{Center for Medical Biotechnology, University of Duisburg-Essen, Germany}

\affiliation{Center for Computational Sciences and Simulation, University of Duisburg-Essen,
Germany}

\affiliation{Center for Water and Environmental Research, University of Duisburg-Essen,
Germany}
\begin{abstract}
We introduce an Interaction and Trade-off based Eco-Evolutionary Model
(ITEEM), in which species are competing for common resources in a
well-mixed system, and their evolution in interaction trait space
is subject to a life-history trade-off between replication rate and
competitive ability. We demonstrate that the strength of the trade-off
has a fundamental impact on eco-evolutionary dynamics, as it imposes
four phases of diversity, including a sharp phase transition. Despite
its minimalism, ITEEM produces without further \emph{ad hoc} features
a remarkable range of observed patterns of eco-evolutionary dynamics.
Most notably we find self-organization towards structured communities
with high and sustainable diversity, in which competing species form
interaction cycles similar to rock-paper-scissors games.
\end{abstract}
\maketitle
Our intuition separates the time scales of fast ecological and slow
evolutionary dynamics, perhaps because we experience the former but
not the latter. However, there is increasing experimental evidence
that this intuition is wrong \cite{Messer2016,Hendry2016,Carroll2007}.
This insight is challenging both ecological and evolutionary theory,
but has also sparked efforts towards unified eco-evolutionary theories
\cite{Carroll2007,Fussmann2007,Schoener2011,Ferriere2013,Moya2014,Hendry2016}.
Here we contribute a new, minimalist model to these efforts, the \emph{Interaction
and Trade-off based Eco-Evolutionary Model} (\emph{ITEEM}). 

\textcolor{black}{The first key idea underlying ITEEM is that }\textcolor{black}{\emph{interactions}}\textcolor{black}{{}
between organisms, mainly competitive interactions, are central to
ecology and to evolution \cite{Allesina2011,Barraclough2015,Weber2017,Coyte2015}.
This insight has inspired work on interaction network topology \cite{Knebel2013,Tang2014,Melo2014,Laird2015,Coyte2015},
and on how these networks evolve and shape diversity \cite{Ginzburg1988,Sole2002a,Tokita2003,Drossel2001,Loeuille2009}.
The second key component of the model is a }\textcolor{black}{\emph{trade-off
}}\textcolor{black}{between interaction traits and replication rate:
better competitors replicate less. Such trade-offs, probably rooted
in differences of energy allocation between life-history traits, have
been observed across biology \cite{Stearns1989,Kneitel2004,Agrawal2010,Maharjan2013,Ferenci2016},
and they were found to be important for emergence and stability of
diversity \cite{Rees1993,Bonsall2004,DeMazancourt2004,Ferenci2016}.}

We show here that ITEEM dynamics closely resembles observed eco-evolutionary
dynamics, such as sympatric speciation \cite{Drossel2000,Coyne2007,Bolnick2007,Herron2013},
emergence of two or more levels of differentiation similar to phylogenetic
structures \cite{Barraclough2003}, large and complex biodiversity
over long times \cite{Herron2013,Kvitek2013}, evolutionary collapses
and extinctions \cite{Karenlampi2014,Sole2002a}, and emergence of
cycles in interaction networks that facilitate species diversification
and coexistence \cite{Mathiesen2011,Bagrow2013,Allesina2011,Laird2015}.
Interestingly, the model shows a unimodal (``humpback'') behavior
of diversity as function of trade-off, with a critical trade-off at
which biodiversity undergoes a phase transition, a behavior observed
in nature \cite{Smith2007,Vallina2014,Nathan2015}.

ITEEM has $N_{s}$ sites of undefined spatial arrangement (well-mixed
system), each providing resources for one organism. We start an eco-evolutionary
simulation with individuals of a single strain occupying a fraction
of the $N_{s}$ sites. Note that in the following we use the term
\emph{strain} for a set of individuals with identical traits. In contrast,
a \emph{species} is a cluster of strains with some diversity (cluster
algorithm described in Supplemental Material SM-1 \cite{SM}).

At every generation or time step $t$, we try $N_{ind}(t)$ (number
of individuals) replications of randomly selected individuals. Each
selected individual of a strain $\alpha$ can replicate with rate
$r_{\alpha}$, with its offspring randomly mutated with rate $\mu$
to new strain $\alpha'$. An individual will vanish if it has reached
its lifespan, drawn at birth from a Poisson distri\textcolor{black}{bution
with overall fixed mean lifespan $\lambda$. }

\textcolor{black}{Each newborn individual is assigned to a randomly
selected site. If the site is empty, the new individual will occupy
it. If the site is already occupied, the new individual competes with
the current holder in a life-or-death struggle; }In that case, the
surviving individual is determined probabilistically by the ``interaction''
$I_{\alpha\beta}$, defined for each pair of strains $\alpha$, $\beta$.
$I_{\alpha\beta}$ is the survival probability of an $\alpha$ individual
in a competitive encounter with a $\beta$ individual, with $I_{\alpha\beta}\in[0,1]$
and $I_{\alpha\beta}+I_{\beta\alpha}=1$.

All interactions $I_{\alpha\beta}$ form an interaction matrix $\mathbf{I}(t)$
that encodes the outcomes of all possible competitive encounters.
If strain $\alpha$ goes extinct, the $\alpha$th row and column of
$\mathbf{I}$ are deleted. Conversely, if a mutation of $\alpha$
generates a new strain $\alpha'$, $\mathbf{I}$ grows by one row
and column:

\begin{eqnarray}
I_{\alpha'\beta} & = & I_{\alpha\beta}+\eta_{\alpha'\beta}\,\,,\nonumber \\
I_{\beta\alpha'} & = & 1-I_{\alpha'\beta}\,\,,\nonumber \\
I_{\alpha'\alpha'} & = & \frac{1}{2}\,\,,\label{eq:Ialphaprimebeta}
\end{eqnarray}

where $\alpha'$ inherits interactions from $\alpha$, but with small
random modification $\eta_{\alpha'\beta}$, drawn from a zero-centered
normal distribution of fixed width $m$. Row$\alpha$ of $\mathbf{I}$
can be considered the ``interaction trait'' $\mathbf{T}_{\alpha}=\left(I_{\alpha1},I_{\alpha2},\ldots,I_{\alpha N_{sp}(t)}\right)$
of strain $\alpha$, with $N_{sp}(t)$ the number of strains at time
$t$. Evolutionary variation of mutants in ITEEM can represent any
phenotypic variation which influences direct interaction of species
and their relative competitive abilities \cite{Thompson1998,Thorpe2011,Bergstrom2015,Thompson1999}.

To implement trade-offs between fecundity and competitive ability,
we introduce a relation between replication rate $r_{\alpha}$ (for
fecundity) and competitive ability $C$, defined as average interaction

\begin{equation}
C(\mathbf{T}_{\alpha})=\frac{1}{N_{sp}(t)-1}\sum_{\beta\neq\alpha}I_{\alpha\beta},\label{eq:def_competitive_ability_C}
\end{equation}

and we let this relation vary with trade-off parameter $s$: 

\begin{equation}
r(\mathbf{T}_{\alpha})=\left(1-C(\mathbf{T}_{\alpha})^{1/s}\right)^{s}.\label{eq:tradeoff}
\end{equation}

With Eq. \ref{eq:tradeoff} better competitive ability leads to lower
fecundity and vice versa. Of course, other functional forms are conceivable.
To systematically study effects of trade-off on dynamics we varied
$s=-\log_{2}(1-\delta)$ with \emph{trade-off strength} $\delta$
covering $[0,1]$ in equidistant steps (SM-2). The larger $\delta$,
the stronger the trade-off. $\delta=0$ makes $r=1$ and thus independent
of $C$.

We compare ITEEM results to the corresponding results of a neutral
model \cite{Hubbell2001}, where we have formally evolving vectors
$\mathbf{T}_{\alpha}$, but fixed and uniform replication rates and
interactions. Accordingly, the neutral model has no trade-off.

ITEEM belongs to the well-established class of generalized Lotka-Volterra
models in the sense that the mean-field version of our stochastic,
agent-based model leads to competitive Lotka-Volterra equations (SM-3).

\paragraph{Generation of diversity}

\begin{center}
\begin{figure*}[t]
\begin{centering}
\includegraphics[width=17cm]{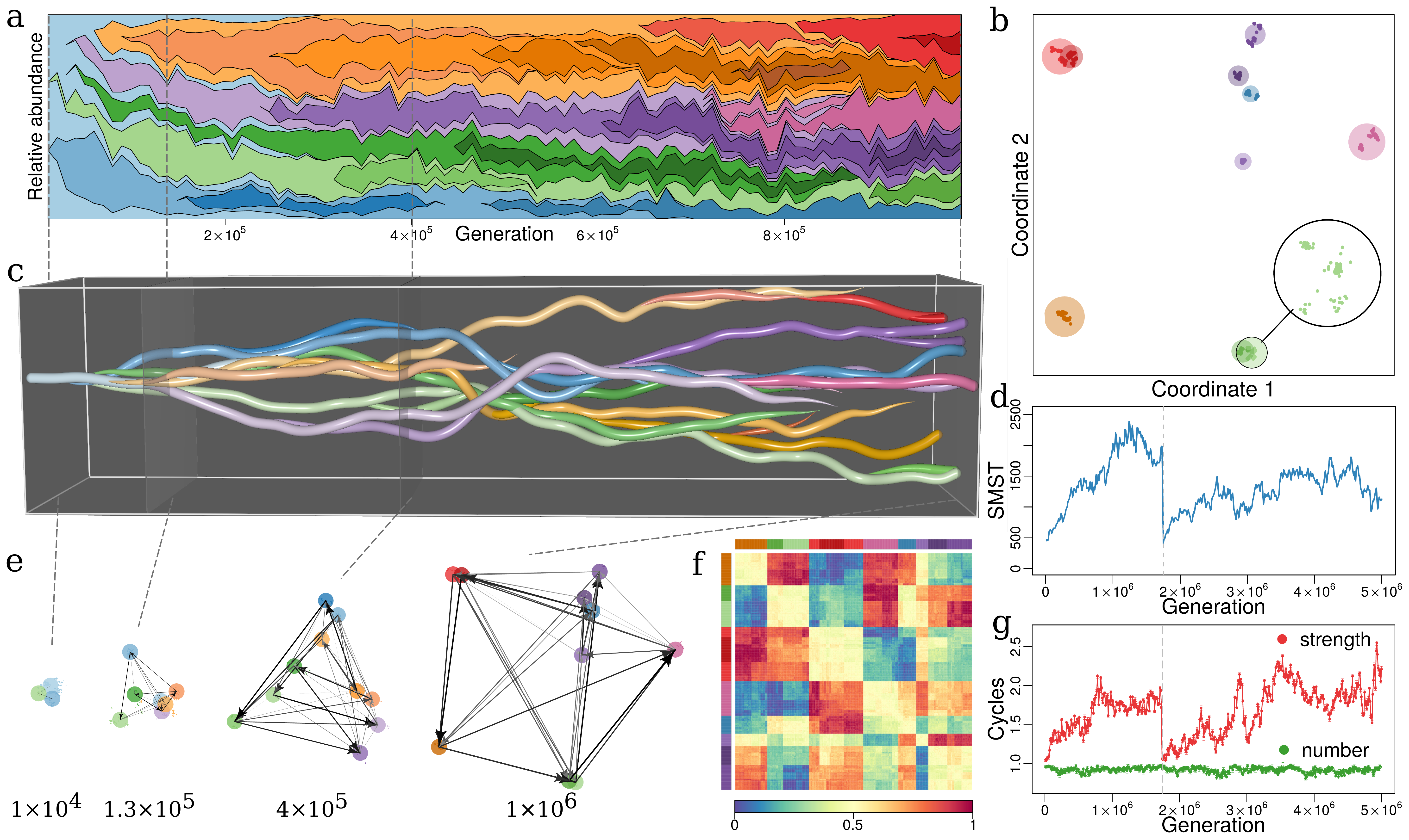}
\end{centering}
\caption{\label{fig:Evolutionary-dynamics}Evolutionary dynamics of a community
driven by competitive interactions, with trade-off between fecundity
and competitive abilities ($\delta=0.5$, $\lambda=300$, $\mu=0.001$,
$m=0.02$, $N_{s}=10^{5}$). (a) \textbf{Species' frequencies over
time} (Muller plot): one color per species, vertical width of each
colored region is relative abundance of respective species. Frequencies
are recorded every $10^{4}$ generations over $10^{6}$ generations.
(b) \textbf{Distribution over trait space}: Snapshot of distribution
in trait space after $10^{6}$ generations, reduced to two dimensions
that explain most of the variance in trait space (SM-4). Points and
discs are strains and species, respectively. Magnified disc in lower
right corner shows strains in the light green species disc. Disc diameter
scales with abundance of species. This snapshot consists of 660 strains
in 10 species. (c) \textbf{Evolutionary dynamics in trait space}:
Snapshots as in panel (b), but concatenated for all times (horizontal
axis), from the monomorphic first generation to $10^{6}$ generations.
(d) \textbf{Functional diversity over time} in terms of the size of
minimum spanning tree (SMST) in trait space (SM-6). At $1.75\times10^{6}$
generations an evolutionary collapse happens in which all species
but one go extinct (vertical dashed line). (e) \textbf{Evolution of
interaction network}: several snapshots from panel (c) with interactions
between species (colored discs) as directed edges. Directions and
strengths of edges given by signs and absolute values, respectively,
of averages over $\bm{I}_{\alpha\beta}-\bm{I}_{\beta\alpha}$, with
$\alpha,\beta$ the component strains of the species linked by edge.
(f) \textbf{Heatmap of interaction matrix }$\mathbf{I}$ for generation
$10^{6}$. Row and column order reflects species clusters, consistent
with panel (b) and indicated by color bars along top and left. Colors
inside heatmap represent interaction rates (color-key along bottom).
(g)\textbf{ Numbers and average strengths of cycles over time} in
green and red, respectively. The strength of a cycle is defined by
its weakest edge. Number and average strength given in units of number
and average strength of equivalent random network, respectively (SM-7).
Right ends in (a) and (c) correspond to panel (b) and (f). Colors
of species are the same in panels (a), (b), (c), (e) and (f). Note
that time scales differ between panels (a), (c) and (d), (g).}
\end{figure*}
\end{center}

Our first question was whether ITEEM is able to generate and sustain
diversity. Since we have a well-mixed system with initially only one
strain, a positive answer implies sympatric diversification, i.e.~the
evolution of new strains and species without geographic isolation
or resource partitioning. In fact, we observe in ITEEM evolution of
new, distinct species, and emergence of sustainable high diversity
(Fig. \ref{fig:Evolutionary-dynamics}a). Remarkably, the emerging
diversity has a clear hierarchical cluster structure (Fig. \ref{fig:Evolutionary-dynamics}b):
at the highest level we see well-separated clusters in trait space
similar to biological \emph{species}. Within these clusters there
are sub-clusters of individual strains (SM-4) \cite{Barraclough2003}.
Both levels of diversity can be quantitatively identified as levels
in the distribution of branch lengths in minimum spanning trees in
trait space (SM-5). This hierarchical diversity is reminiscent of
the phylogenetic structures in biology \cite{Barraclough2003}. Overall,
the model shows evolutionary divergence from one strain to several
species consisting of a total of hundreds of co-existing strains over
millions of generations (Fig. \ref{fig:Evolutionary-dynamics}c, and
SM-6.1). Depending on trade-off parameter $\delta$, this high diversity
is often sustainable over hundreds of thousands of generations. Collapses
to low diversity occur rarely and are usually followed by recovery
of diversity (Fig. \ref{fig:Evolutionary-dynamics}d, and SM-6.1).

The observed divergence contradicts the long-held view of sequential
fixation in asexual populations \cite{Muller1932}. Instead, we see
frequently concurrent speciation with emergence of two or more species
in quick succession (Fig. \ref{fig:Evolutionary-dynamics}a), in agreement
with recent results from long-term bacterial cultures \cite{Herron2013,Maddamsetti2015,Kvitek2013}. 

Our model allows to study speciation in detail, e.g.~in terms of
interaction network dynamics. The interaction matrix $\mathbf{I}$
defines a complete graph, and we determined direction and strength
of interaction edges between two strains $\alpha,\beta$ as sign and
size of $I_{\alpha\beta}-I_{\beta\alpha}$. Accordingly, for the interaction
network of \emph{species }(i.e.~clusters of strains) we computed
directed edges between any two species by averaging over inter-cluster
edges between the strains in these clusters (Fig. \ref{fig:Evolutionary-dynamics}e).
Three or more directed edges can form cycles of strains in which each
strain competes successfully against one cycle neighbor but loses
against the other neighbor, a configuration corresponding to rock-paper-scissors
games \cite{Szolnoki2014}. Such intransitive interactions have been
observed in nature \cite{Sinervo1996,Lankau2007,Bergstrom2015}, and
it has been shown that they stabilize a system driven by competitive
interactions \cite{Mitarai2012,Allesina2011,Mathiesen2011}. In fact,
we find that the increase of diversity as measured by e.g.~richness,
entropy, or functional diversity (SM-6), coincides with growth of
average cycle strength (Figs \ref{fig:Evolutionary-dynamics}d, g
and SM-7).

\noindent 

\paragraph{Impact of trade-off and lifespan on diversity}

\begin{center}
\begin{figure}
\begin{centering}
\includegraphics[width=8.5cm]{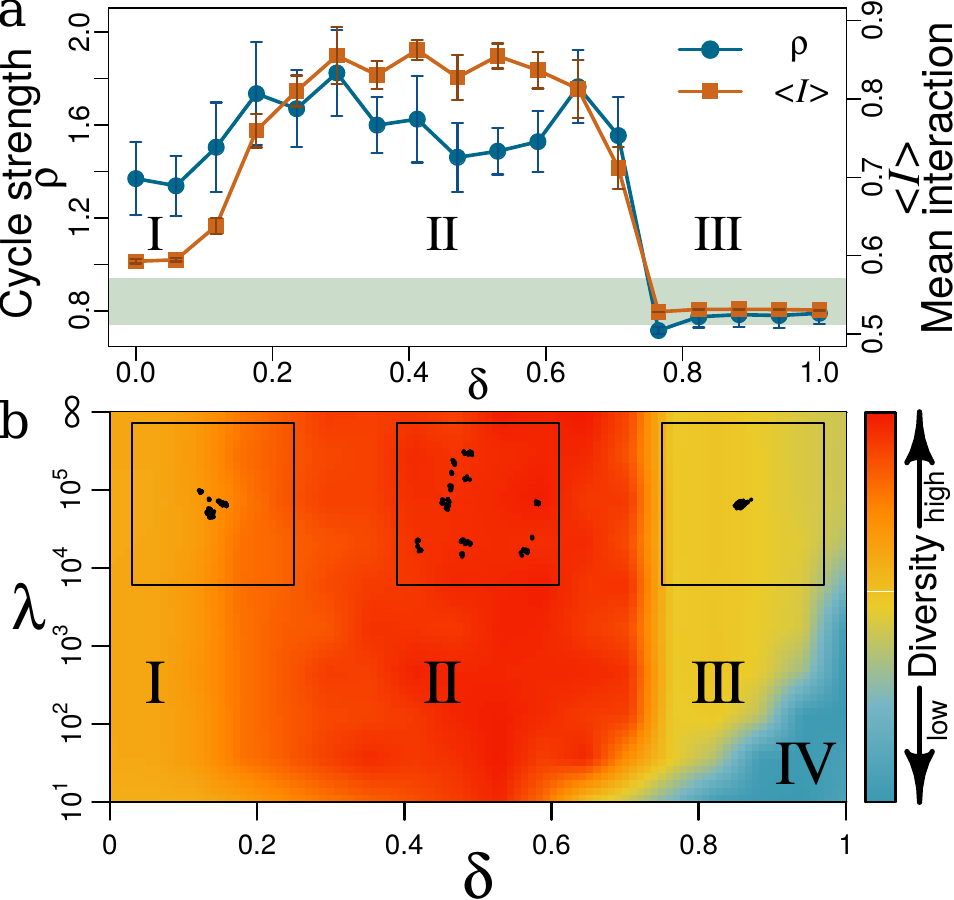}
\end{centering}
\caption{\label{fig:Phase-diagram} Effects of trade-off $\delta$ and lifespan
$\lambda$ on community structure and diversity. (a) Average interaction
rates $\left\langle I\right\rangle $ (orange squares) and average
strength of cycles $\rho$ (blue circles) as function of $\delta$.
Average cycle strength is given in units of average strength of random
networks for the respective trade-off (SM-7). Averages are calculated
over three different simulations, each over $5\times10^{6}$generations
with $\mu=0.001$, $m=0.02$, $\lambda=\infty$ and $N_{S}=10^{5}$.
Error bars are standard deviations averaged over three concatenated
simulations. The shaded area marks cycle strength for a neutral model
with corresponding parameters $\pm$ standard deviation. (b) Phase
diagram of diversity as function of $\delta$ and $\lambda$. Diversity
is given as consensus of several quantities (SM-6). Four phases (I-IV)
can be distinguished. Insets at the top margin are representative
MDS plots (SM-4) of strain distributions in trait space, as in Fig.
\ref{fig:Evolutionary-dynamics}b, with $\lambda=10^{5}$ but different
values of $\delta$ (left to right: I with $\delta=0.11$; II with
$\delta=0.5$; III with $\delta=0.89$). Panel (a) corresponds to
a horizontal cross-section through the phase diagram in panel (b)
with $\lambda=\infty$ for $\left\langle I\right\rangle $ and $\rho$
as diversity measures.}
\end{figure}
\end{center}

The eco-evolutionary dynamics described above depend on lifespan and
trade-off between replication and competitive ability. To show this
we study properties of interaction matrix and trait diversity. Fig.
\ref{fig:Phase-diagram} relates average interaction rate $\left\langle I\right\rangle $
and average cycle strength $\rho$ to trade-off parameter $\delta$
at fixed lifespan $\lambda$. Fig. \ref{fig:Phase-diagram}b summarizes
the behavior of diversity as function of $\delta$ and $\lambda$.
Overall, we see in this phase diagram a weak dependency on $\lambda$
and a strong impact of $\delta$, with four distinct phases (I-IV)
from low to high $\delta$.

Without trade-off, strains do not have to sacrifice replication rate
for better competitive abilities. We have a low-diversity population
dominated by Darwinian demons, species with high competitive ability
and replication rate. Quick predominance of such strategies impedes
formation of a diverse network. Increasing $\delta$ in phase I ($\delta\lesssim0.2$)
slightly increases $\left\langle I\right\rangle $ and $\rho$ (Fig.
\ref{fig:Phase-diagram}a): biotic selection pressure exerted by inter-species
interactions starts to generate diverse communities (left inset in
Fig. \ref{fig:Phase-diagram}b, SM-6). However, the weak trade-off
still favors investing in higher competitive ability. When increasing
$\delta$ further (phase II), trade-off starts to force strains to
choose between higher replication rate $r$ or better competitive
abilities $C$. Neither extreme generates viable species: sacrificing
$r$ completely for maximum $C$ stalls species dynamics, whereas
maximum $r$ leads to inferior $C$. Thus strains seek middle ground
values in both $r$ and $C$. The nature of $C$ as mean of interactions
(Eq. \ref{eq:def_competitive_ability_C}) allows for many combinations
of interaction traits with approximately the same mean. Thus in a
middle range of $r$ and $C$, many strategies with the same overall
fitness are possible, which is a condition of diversity. From this
multitude of strategies, sets of trait combinations emerge in which
strains with different combinations keep each other in check, e.g.~in
the form of competitive rock-paper-scissors-like cycles between species
described above. An equivalent interpretation is the emergence of
diverse sets of non-overlapping compartments or trait space niches
(Fig. \ref{fig:Evolutionary-dynamics}b,f). Diversity in this phase
II is the highest and most stable (middle inset in Fig. \ref{fig:Phase-diagram}b,
SM-6). As $\delta$ approaches $0.7$, $\left\langle I\right\rangle $
and $\rho$ plummet (Fig. \ref{fig:Phase-diagram}a) to interaction
rates comparable to noise level $m$, and a cycle strength typical
for the neutral model (horizontal gray ribbon in Fig. \ref{fig:Phase-diagram}a),
respectively. The sharp drop of $\left\langle I\right\rangle $ and
$\rho$ at $\delta\approx0.7$ is reminiscent of a phase transition.
As expected for a phase transition, the steepness increases with system
size (SM-8). For $\delta\gtrsim0.7$ interaction rates never grow
and no structure emerges; diversity remains low and close to a neutral
system. The sharp transition at $\delta\approx0.7$ which is visible
in practically all diversity measures (between phases II and III in
Fig. \ref{fig:Phase-diagram}b, SM-6) is a transition from a system
dominated by biotic selection pressure to a neutral system. In high-trade-off
phase III, any small change in $C$ changes $r$ drastically. For
instance, given a strain $S$ with $r$ and $C$, a closely related
mutant $S'$ with $C'\lessapprox C$ will have $r'\gg r$ (because
of the large trade-off), and therefore will invade $S$ quickly. Thus,
diversity in phase III will remain stable and low, characterized by
a group of similar strains with no effective interaction and hence
no diversification to distinct species (right inset in Fig. \ref{fig:Phase-diagram}b,
SM-6).

In this high trade-off regime, lifespan comes into play: here, decreasing
$\lambda$ can make lives too short for replication. These hostile
conditions minimize diversity and favor extinction (phase IV).

\noindent 

\paragraph{Trade-off, resource availability, and diversity}

There is a well-known but not well understood unimodal relationship
between biomass productivity and diversity (``humpback curve'',\cite{Smith2007,Vallina2014}):
diversity culminates once at middle values of productivity. This behavior
is reminiscent of horizontal sections through the phase diagram in
Fig. \ref{fig:Phase-diagram}b, though here the driving parameter
is not productivity but trade-off. However, we can make the following
argument for a monotonous relation between productivity and trade-off.
First we note that biomass productivity is a function of available
resources: the larger the available resources, the higher the productivity.
This allows us to argue in terms of available resources. If then a
species has a high replication rate in an environment with scarce
resources, its individuals will not be very competitive since for
each of the numerous offspring individuals there is little material
or energy available. On the other hand, if a species under these resource-limited
conditions has competitively constructed individuals it cannot produce
many of them. This corresponds to a strong trade-off between replication
and competitive ability for scarce resources. At the opposite, rich
end of the resource scale, species are not confronted with such hard
choices between replication rate and competitive ability, i.e.~we
have a weak trade-off. Taken together, the trade-off axis should roughly
correspond to the inverted resource axis: strong trade-off for poor
resources (or low productivity), weak trade-off for rich resources
(or high productivity); a detailed analytical derivation will be presented
elsewhere. The fact that ITEEM produces this humpback curve that is
frequently observed in planktonic systems \cite{Vallina2014} proposes
trade-off as underlying mechanism of this productivity-diversity relation.

\paragraph{Frequency-dependent selection }

Observation of eco-evolutionary trajectories as in Fig. \ref{fig:Evolutionary-dynamics}
suggested the hypothesis that speciation events in ITEEM simulations
do not occur with a constant rate and independently of each other,
but that one speciation makes a following speciation more likely.
We therefore tested whether the distribution of time between speciation
or extinction events is compatible with a constant rate Poisson process
(SM-9). At long inter-event times we see the same decaying distribution
for the Poisson process and for the ITEEM data. However, for shorter
times there are significant deviations from a Poisson process for
speciation and extinction events: at inter-event times of around $10\text{\textsuperscript{4}}$
the number of events \emph{decreases }for a Poisson process but \emph{increases
}in ITEEM simulations. This confirms the above hypothesis that new
species increase the probability for generation of further species,
and additionally that loss of a species makes further losses more
likely. This result is similar to the frequency-dependent selection
observed in microbial systems where new species open new niches for
further species, or the loss of species causes the loss of dependent
species \cite{Herron2013,Maddamsetti2015}.

\paragraph{Effect of mutation rate on diversity}

Simulations with different mutation rates ($\mu=10^{-4},5\times10^{-4},10^{-3},5\times10^{-3}$)
show that in ITEEM diversity grows faster and to a higher level with
increasing mutation rate, but without changing the overall structure
of the phase diagram (SM-10). One interesting tendency is that for
higher mutation rates, the lifespan becomes more important at the
interface of regions III and IV (high trade-offs), leading to an expansion
of region III at the expense of hostile region IV: long lifespans
in combination with high mutation rate establish low but viable diversity
at strong trade-offs.

\paragraph{Comparison of ITEEM with neutral model}

The neutral model introduced in the Model section has no meaningful
interaction traits, and consequently no meaningful competitive ability
or trade-off with replication rate. Instead, it evolves solely by
random drift in phenotype space. Similarly to ITEEM, the neutral model
generates a clumpy structure in trait space (SM-11), though here the
species clusters are much closer and thus the functional diversity
much lower. This can be demonstrated quantitatively by the size of
the minimum spanning tree of populations in trait space that are much
smaller and much less dynamic for the neutral model than for ITEEM
at moderate trade-off (SM-11). For high trade-offs (region III, Fig.
\ref{fig:Phase-diagram}b), diversity and number of strong cycles
in ITEEM are comparable to the neutral model (Fig. \ref{fig:Phase-diagram}a).

Interaction based eco-evolutionary models have received some attention
in the past \cite{Ginzburg1988,Sole2002a,Tokita2003,Shtilerman2015}
but then were almost forgotten, despite remarkable results. We think
that these works have pointed to a possible solution of a hard problem:
The complexity of evolving ecosystems is immense, and it is therefore
difficult to find a representation suitable for the development of
a statistical mechanics that enables qualitative and quantitative
analysis \cite{Weber2017}. Modeling in terms of interaction traits,
rather than detailed descriptions of genotypes or phenotypes, then
coarse-grains these complex systems in a natural, biologically meaningful
way. 

Despite these advantages, interaction based models so far have not
shown some key features of real systems, e.g. emergence of large,
stable and complex diversity, or mass extinctions with the subsequent
recovery of diversity \cite{Tokita2003,Karenlampi2014}. Therefore,
interaction based models were supplemented by \emph{ad hoc }features,
such as special types of mutations \cite{Tokita2003}, induced extinctions
\cite{Vandewalle1995}, or enforcement of partially connected interaction
graphs \cite{Karenlampi2014}. 

Trade-off between replication and competitive ability have now been
experimentally established as essential to living systems \cite{Stearns1989,Agrawal2010}.
Our results with ITEEM show that trade-offs fundamentally impact eco-evolutionary
dynamics, in agreement with other eco-evolutionary models with trade-off
\cite{Huisman2001,Bonsall2004,DeMazancourt2004,Beardmore2011}. Remarkably,
we observe with ITEEM sustained high diversity in a well-mixed homogeneous
system, without violating the competitive exclusion principle. This
is possible because moderate life-history trade-offs force evolving
species to adopt different strategies or, in other words, lead to
the emergence of well-separated niches in interaction space.

The current model has important limitations. For instance, the trade-off
formulation was chosen to reflect reasonable properties in a minimalistic
way, that should be revised or refined as more experimental data become
available. Secondly, we have assumed a single, limiting resource in
a well-mixed system to investigate the mechanisms behind diversification
in competitive communities and possibility of niche differentiation
without resource partitioning or geographic isolation. However, in
nature, there will in general be several limiting resources and abiotic
factors. It is possible to include those as additional rows and columns
in the interaction matrix $\mathbf{I}$.

Despite its simplifications, ITEEM reproduces in a single framework
several phenomena of eco-evolutionary dynamics that previously were
addressed with a range of distinct models or not at all, namely sympatric
and concurrent speciation with the emergence of new niches in the
community, recovery after mass-extinctions, large and sustained functional
diversity with hierarchical organization, and a unimodal diversity
distribution as function of trade-off between replication and competition.
The model allows detailed analysis of mechanisms and could guide experimental
tests. 
\begin{acknowledgments}
We thank S. Moghimi-Araghi for helpful suggestions on trade-off function.
\end{acknowledgments}
\bibliographystyle{apsrev}

\begin{thebibliography}{56}
\expandafter\ifx\csname natexlab\endcsname\relax\def\natexlab#1{#1}\fi
\expandafter\ifx\csname bibnamefont\endcsname\relax
  \def\bibnamefont#1{#1}\fi
\expandafter\ifx\csname bibfnamefont\endcsname\relax
  \def\bibfnamefont#1{#1}\fi
\expandafter\ifx\csname citenamefont\endcsname\relax
  \def\citenamefont#1{#1}\fi
\expandafter\ifx\csname url\endcsname\relax
  \def\url#1{\texttt{#1}}\fi
\expandafter\ifx\csname urlprefix\endcsname\relax\def\urlprefix{URL }\fi
\providecommand{\bibinfo}[2]{#2}
\providecommand{\eprint}[2][]{\url{#2}}

\bibitem[{\citenamefont{Messer et~al.}(2016)\citenamefont{Messer, Ellner, and
  Hairston}}]{Messer2016}
\bibinfo{author}{\bibfnamefont{P.~W.} \bibnamefont{Messer}},
  \bibinfo{author}{\bibfnamefont{S.~P.} \bibnamefont{Ellner}},
  \bibnamefont{and} \bibinfo{author}{\bibfnamefont{N.~G.}
  \bibnamefont{Hairston}}, \bibinfo{journal}{Trends Genet.}
  \textbf{\bibinfo{volume}{32}}, \bibinfo{pages}{408} (\bibinfo{year}{2016}).

\bibitem[{\citenamefont{Hendry}(2016)}]{Hendry2016}
\bibinfo{author}{\bibfnamefont{A.~P.} \bibnamefont{Hendry}},
  \emph{\bibinfo{title}{Eco-evolutionary Dynamics}}
  (\bibinfo{publisher}{Princeton University Press},
  \bibinfo{address}{Princeton}, \bibinfo{year}{2016}).

\bibitem[{\citenamefont{Carrol et~al.}(2007)\citenamefont{Carrol, Hendry,
  Reznick, and Fox}}]{Carroll2007}
\bibinfo{author}{\bibfnamefont{S.~P.} \bibnamefont{Carrol}},
  \bibinfo{author}{\bibfnamefont{A.~P.} \bibnamefont{Hendry}},
  \bibinfo{author}{\bibfnamefont{D.~N.} \bibnamefont{Reznick}},
  \bibnamefont{and} \bibinfo{author}{\bibfnamefont{C.~W.} \bibnamefont{Fox}},
  \bibinfo{journal}{Funct. Ecol.} \textbf{\bibinfo{volume}{21}},
  \bibinfo{pages}{387} (\bibinfo{year}{2007}).

\bibitem[{\citenamefont{Fussmann et~al.}(2007)\citenamefont{Fussmann, Loreau,
  and Abrams}}]{Fussmann2007}
\bibinfo{author}{\bibfnamefont{G.~F.} \bibnamefont{Fussmann}},
  \bibinfo{author}{\bibfnamefont{M.}~\bibnamefont{Loreau}}, \bibnamefont{and}
  \bibinfo{author}{\bibfnamefont{P.~A.} \bibnamefont{Abrams}},
  \bibinfo{journal}{Funct. Ecol.} \textbf{\bibinfo{volume}{21}},
  \bibinfo{pages}{465} (\bibinfo{year}{2007}).

\bibitem[{\citenamefont{Schoener}(2011)}]{Schoener2011}
\bibinfo{author}{\bibfnamefont{T.~W.} \bibnamefont{Schoener}},
  \bibinfo{journal}{Science (80-. ).} \textbf{\bibinfo{volume}{331}},
  \bibinfo{pages}{426} (\bibinfo{year}{2011}).

\bibitem[{\citenamefont{Ferriere and Legendre}(2012)}]{Ferriere2013}
\bibinfo{author}{\bibfnamefont{R.}~\bibnamefont{Ferriere}} \bibnamefont{and}
  \bibinfo{author}{\bibfnamefont{S.}~\bibnamefont{Legendre}},
  \bibinfo{journal}{Philos. Trans. R. Soc. B Biol. Sci.}
  \textbf{\bibinfo{volume}{368}}, \bibinfo{pages}{20120081}
  (\bibinfo{year}{2012}).

\bibitem[{\citenamefont{Moya-Lara{\~n}o
  et~al.}(2014)\citenamefont{Moya-Lara{\~n}o, Rowntree, and
  Woodward}}]{Moya2014}
\bibinfo{author}{\bibfnamefont{J.}~\bibnamefont{Moya-Lara{\~n}o}},
  \bibinfo{author}{\bibfnamefont{J.}~\bibnamefont{Rowntree}}, \bibnamefont{and}
  \bibinfo{author}{\bibfnamefont{G.}~\bibnamefont{Woodward}},
  \emph{\bibinfo{title}{Advances in Ecological Research: Eco-evolutionary
  dynamics}}, vol.~\bibinfo{volume}{50} (\bibinfo{publisher}{Academic Press},
  \bibinfo{year}{2014}).

\bibitem[{\citenamefont{Allesina and Levine}(2011)}]{Allesina2011}
\bibinfo{author}{\bibfnamefont{S.}~\bibnamefont{Allesina}} \bibnamefont{and}
  \bibinfo{author}{\bibfnamefont{J.~M.} \bibnamefont{Levine}},
  \bibinfo{journal}{Proc. Natl. Acad. Sci.} \textbf{\bibinfo{volume}{108}},
  \bibinfo{pages}{5638} (\bibinfo{year}{2011}).

\bibitem[{\citenamefont{Barraclough}(2015)}]{Barraclough2015}
\bibinfo{author}{\bibfnamefont{T.~G.} \bibnamefont{Barraclough}},
  \bibinfo{journal}{Annu. Rev. Ecol. Evol. Syst.}
  \textbf{\bibinfo{volume}{46}}, \bibinfo{pages}{25} (\bibinfo{year}{2015}).

\bibitem[{\citenamefont{Weber et~al.}(2017)\citenamefont{Weber, Wagner, Best,
  Harmon, and Matthews}}]{Weber2017}
\bibinfo{author}{\bibfnamefont{M.~G.} \bibnamefont{Weber}},
  \bibinfo{author}{\bibfnamefont{C.~E.} \bibnamefont{Wagner}},
  \bibinfo{author}{\bibfnamefont{R.~J.} \bibnamefont{Best}},
  \bibinfo{author}{\bibfnamefont{L.~J.} \bibnamefont{Harmon}},
  \bibnamefont{and} \bibinfo{author}{\bibfnamefont{B.}~\bibnamefont{Matthews}},
  \bibinfo{journal}{Trends Ecol. Evol.} \textbf{\bibinfo{volume}{32}},
  \bibinfo{pages}{291} (\bibinfo{year}{2017}).

\bibitem[{\citenamefont{Coyte et~al.}(2015)\citenamefont{Coyte, Schluter, and
  Foster}}]{Coyte2015}
\bibinfo{author}{\bibfnamefont{K.~Z.} \bibnamefont{Coyte}},
  \bibinfo{author}{\bibfnamefont{J.}~\bibnamefont{Schluter}}, \bibnamefont{and}
  \bibinfo{author}{\bibfnamefont{K.~R.} \bibnamefont{Foster}},
  \bibinfo{journal}{Science (80-. ).} \textbf{\bibinfo{volume}{350}},
  \bibinfo{pages}{663} (\bibinfo{year}{2015}).

\bibitem[{\citenamefont{Knebel et~al.}(2013)\citenamefont{Knebel, Kr{\"{u}}ger,
  Weber, and Frey}}]{Knebel2013}
\bibinfo{author}{\bibfnamefont{J.}~\bibnamefont{Knebel}},
  \bibinfo{author}{\bibfnamefont{T.}~\bibnamefont{Kr{\"{u}}ger}},
  \bibinfo{author}{\bibfnamefont{M.~F.} \bibnamefont{Weber}}, \bibnamefont{and}
  \bibinfo{author}{\bibfnamefont{E.}~\bibnamefont{Frey}},
  \bibinfo{journal}{Phys. Rev. Lett.} \textbf{\bibinfo{volume}{110}},
  \bibinfo{pages}{168106} (\bibinfo{year}{2013}).

\bibitem[{\citenamefont{Tang et~al.}(2014)\citenamefont{Tang, Pawar, and
  Allesina}}]{Tang2014}
\bibinfo{author}{\bibfnamefont{S.}~\bibnamefont{Tang}},
  \bibinfo{author}{\bibfnamefont{S.}~\bibnamefont{Pawar}}, \bibnamefont{and}
  \bibinfo{author}{\bibfnamefont{S.}~\bibnamefont{Allesina}},
  \bibinfo{journal}{Ecol. Lett.} \textbf{\bibinfo{volume}{17}},
  \bibinfo{pages}{1094} (\bibinfo{year}{2014}).

\bibitem[{\citenamefont{Melo and Marroig}(2015)}]{Melo2014}
\bibinfo{author}{\bibfnamefont{D.}~\bibnamefont{Melo}} \bibnamefont{and}
  \bibinfo{author}{\bibfnamefont{G.}~\bibnamefont{Marroig}},
  \bibinfo{journal}{Proc. Natl. Acad. Sci.} \textbf{\bibinfo{volume}{112}},
  \bibinfo{pages}{470} (\bibinfo{year}{2015}).

\bibitem[{\citenamefont{Laird and Schamp}(2015)}]{Laird2015}
\bibinfo{author}{\bibfnamefont{R.~A.} \bibnamefont{Laird}} \bibnamefont{and}
  \bibinfo{author}{\bibfnamefont{B.~S.} \bibnamefont{Schamp}},
  \bibinfo{journal}{J. Theor. Biol.} \textbf{\bibinfo{volume}{365}},
  \bibinfo{pages}{149} (\bibinfo{year}{2015}).

\bibitem[{\citenamefont{Ginzburg et~al.}(1988)\citenamefont{Ginzburg,
  Ak{\c{c}}akaya, and Kim}}]{Ginzburg1988}
\bibinfo{author}{\bibfnamefont{L.~R.} \bibnamefont{Ginzburg}},
  \bibinfo{author}{\bibfnamefont{H.~R.} \bibnamefont{Ak{\c{c}}akaya}},
  \bibnamefont{and} \bibinfo{author}{\bibfnamefont{J.}~\bibnamefont{Kim}},
  \bibinfo{journal}{J. Theor. Biol.} \textbf{\bibinfo{volume}{133}},
  \bibinfo{pages}{513} (\bibinfo{year}{1988}).

\bibitem[{\citenamefont{Sol{\'{e}}}(2002)}]{Sole2002a}
\bibinfo{author}{\bibfnamefont{R.~V.} \bibnamefont{Sol{\'{e}}}}, in
  \emph{\bibinfo{booktitle}{Biol. Evol. Stat. Phys.}}
  (\bibinfo{publisher}{Springer Berlin Heidelberg}, \bibinfo{address}{Berlin,
  Heidelberg}, \bibinfo{year}{2002}), pp. \bibinfo{pages}{312--337}.

\bibitem[{\citenamefont{Tokita and Yasutomi}(2003)}]{Tokita2003}
\bibinfo{author}{\bibfnamefont{K.}~\bibnamefont{Tokita}} \bibnamefont{and}
  \bibinfo{author}{\bibfnamefont{A.}~\bibnamefont{Yasutomi}},
  \bibinfo{journal}{Theor. Popul. Biol.} \textbf{\bibinfo{volume}{63}},
  \bibinfo{pages}{131} (\bibinfo{year}{2003}).

\bibitem[{\citenamefont{Drossel et~al.}(2001)\citenamefont{Drossel, Higgs, and
  McKane}}]{Drossel2001}
\bibinfo{author}{\bibfnamefont{B.}~\bibnamefont{Drossel}},
  \bibinfo{author}{\bibfnamefont{P.~G.} \bibnamefont{Higgs}}, \bibnamefont{and}
  \bibinfo{author}{\bibfnamefont{A.~J.} \bibnamefont{McKane}},
  \bibinfo{journal}{J. Theor. Biol.} \textbf{\bibinfo{volume}{208}},
  \bibinfo{pages}{91} (\bibinfo{year}{2001}).

\bibitem[{\citenamefont{Loeuille and Loreau}(2009)}]{Loeuille2009}
\bibinfo{author}{\bibfnamefont{N.}~\bibnamefont{Loeuille}} \bibnamefont{and}
  \bibinfo{author}{\bibfnamefont{M.}~\bibnamefont{Loreau}}, in
  \emph{\bibinfo{booktitle}{Community Ecol.}} (\bibinfo{publisher}{Oxford
  University Press}, \bibinfo{year}{2009}), vol. \bibinfo{volume}{102}, pp.
  \bibinfo{pages}{163--178}.

\bibitem[{\citenamefont{Stearns}(1989)}]{Stearns1989}
\bibinfo{author}{\bibfnamefont{S.~C.} \bibnamefont{Stearns}},
  \bibinfo{journal}{Funct. Ecol.} \textbf{\bibinfo{volume}{3}},
  \bibinfo{pages}{259} (\bibinfo{year}{1989}).

\bibitem[{\citenamefont{Kneitel and Chase}(2004)}]{Kneitel2004}
\bibinfo{author}{\bibfnamefont{J.~M.} \bibnamefont{Kneitel}} \bibnamefont{and}
  \bibinfo{author}{\bibfnamefont{J.~M.} \bibnamefont{Chase}},
  \bibinfo{journal}{Ecol. Lett.} \textbf{\bibinfo{volume}{7}},
  \bibinfo{pages}{69} (\bibinfo{year}{2004}).

\bibitem[{\citenamefont{Agrawal et~al.}(2010)\citenamefont{Agrawal, Conner, and
  Rasmann}}]{Agrawal2010}
\bibinfo{author}{\bibfnamefont{A.~A.} \bibnamefont{Agrawal}},
  \bibinfo{author}{\bibfnamefont{J.~K.} \bibnamefont{Conner}},
  \bibnamefont{and} \bibinfo{author}{\bibfnamefont{S.}~\bibnamefont{Rasmann}},
  in \emph{\bibinfo{booktitle}{Evolution since Darwin: the first 150 years}},
  edited by \bibinfo{editor}{\bibfnamefont{M.~A.} \bibnamefont{Bell}},
  \bibinfo{editor}{\bibfnamefont{D.~J.} \bibnamefont{Futuyama}},
  \bibinfo{editor}{\bibfnamefont{W.~F.} \bibnamefont{Eanes}}, \bibnamefont{and}
  \bibinfo{editor}{\bibfnamefont{J.~S.} \bibnamefont{Levinton}}
  (\bibinfo{publisher}{Sinauer Associates, Inc.}, \bibinfo{address}{Sunderland,
  MA}, \bibinfo{year}{2010}), chap.~\bibinfo{chapter}{10}, pp.
  \bibinfo{pages}{243--268}.

\bibitem[{\citenamefont{Maharjan et~al.}(2013)\citenamefont{Maharjan, Nilsson,
  Sung, Haynes, Beardmore, Hurst, Ferenci, and Gudelj}}]{Maharjan2013}
\bibinfo{author}{\bibfnamefont{R.}~\bibnamefont{Maharjan}},
  \bibinfo{author}{\bibfnamefont{S.}~\bibnamefont{Nilsson}},
  \bibinfo{author}{\bibfnamefont{J.}~\bibnamefont{Sung}},
  \bibinfo{author}{\bibfnamefont{K.}~\bibnamefont{Haynes}},
  \bibinfo{author}{\bibfnamefont{R.~E.} \bibnamefont{Beardmore}},
  \bibinfo{author}{\bibfnamefont{L.~D.} \bibnamefont{Hurst}},
  \bibinfo{author}{\bibfnamefont{T.}~\bibnamefont{Ferenci}}, \bibnamefont{and}
  \bibinfo{author}{\bibfnamefont{I.}~\bibnamefont{Gudelj}},
  \bibinfo{journal}{Ecol. Lett.} \textbf{\bibinfo{volume}{16}},
  \bibinfo{pages}{1267} (\bibinfo{year}{2013}).

\bibitem[{\citenamefont{Ferenci}(2016)}]{Ferenci2016}
\bibinfo{author}{\bibfnamefont{T.}~\bibnamefont{Ferenci}},
  \bibinfo{journal}{Trends Microbiol.} \textbf{\bibinfo{volume}{24}},
  \bibinfo{pages}{209} (\bibinfo{year}{2016}).

\bibitem[{\citenamefont{Rees}(1993)}]{Rees1993}
\bibinfo{author}{\bibfnamefont{M.}~\bibnamefont{Rees}},
  \bibinfo{journal}{Nature} \textbf{\bibinfo{volume}{366}},
  \bibinfo{pages}{150} (\bibinfo{year}{1993}).

\bibitem[{\citenamefont{Bonsall}(2004)}]{Bonsall2004}
\bibinfo{author}{\bibfnamefont{M.~B.} \bibnamefont{Bonsall}},
  \bibinfo{journal}{Science (80-. ).} \textbf{\bibinfo{volume}{306}},
  \bibinfo{pages}{111} (\bibinfo{year}{2004}).

\bibitem[{\citenamefont{de~Mazancourt and Dieckmann}(2004)}]{DeMazancourt2004}
\bibinfo{author}{\bibfnamefont{C.}~\bibnamefont{de~Mazancourt}}
  \bibnamefont{and}
  \bibinfo{author}{\bibfnamefont{U.}~\bibnamefont{Dieckmann}},
  \bibinfo{journal}{Am. Nat.} \textbf{\bibinfo{volume}{164}},
  \bibinfo{pages}{765} (\bibinfo{year}{2004}).

\bibitem[{\citenamefont{Drossel and McKane}(2000)}]{Drossel2000}
\bibinfo{author}{\bibfnamefont{B.}~\bibnamefont{Drossel}} \bibnamefont{and}
  \bibinfo{author}{\bibfnamefont{A.~J.} \bibnamefont{McKane}},
  \bibinfo{journal}{J. Theor. Biol.} \textbf{\bibinfo{volume}{204}},
  \bibinfo{pages}{467} (\bibinfo{year}{2000}).

\bibitem[{\citenamefont{Coyne}(2007)}]{Coyne2007}
\bibinfo{author}{\bibfnamefont{J.~A.} \bibnamefont{Coyne}},
  \bibinfo{journal}{Curr. Biol.} \textbf{\bibinfo{volume}{17}},
  \bibinfo{pages}{R787} (\bibinfo{year}{2007}).

\bibitem[{\citenamefont{Bolnick and Fitzpatrick}(2007)}]{Bolnick2007}
\bibinfo{author}{\bibfnamefont{D.~I.} \bibnamefont{Bolnick}} \bibnamefont{and}
  \bibinfo{author}{\bibfnamefont{B.~M.} \bibnamefont{Fitzpatrick}},
  \bibinfo{journal}{Annu. Rev. Ecol. Evol. Syst.}
  \textbf{\bibinfo{volume}{38}}, \bibinfo{pages}{459} (\bibinfo{year}{2007}).

\bibitem[{\citenamefont{Herron and Doebeli}(2013)}]{Herron2013}
\bibinfo{author}{\bibfnamefont{M.~D.} \bibnamefont{Herron}} \bibnamefont{and}
  \bibinfo{author}{\bibfnamefont{M.}~\bibnamefont{Doebeli}},
  \bibinfo{journal}{PLoS Biol.} \textbf{\bibinfo{volume}{11}},
  \bibinfo{pages}{e1001490} (\bibinfo{year}{2013}).

\bibitem[{\citenamefont{Barraclough et~al.}(2003)\citenamefont{Barraclough,
  Birky, and Burt}}]{Barraclough2003}
\bibinfo{author}{\bibfnamefont{T.~G.} \bibnamefont{Barraclough}},
  \bibinfo{author}{\bibfnamefont{C.~W.} \bibnamefont{Birky}}, \bibnamefont{and}
  \bibinfo{author}{\bibfnamefont{A.}~\bibnamefont{Burt}},
  \bibinfo{journal}{Evolution (N. Y).} \textbf{\bibinfo{volume}{57}},
  \bibinfo{pages}{2166} (\bibinfo{year}{2003}).

\bibitem[{\citenamefont{Kvitek and Sherlock}(2013)}]{Kvitek2013}
\bibinfo{author}{\bibfnamefont{D.~J.} \bibnamefont{Kvitek}} \bibnamefont{and}
  \bibinfo{author}{\bibfnamefont{G.}~\bibnamefont{Sherlock}},
  \bibinfo{journal}{PLoS Genet.} \textbf{\bibinfo{volume}{9}},
  \bibinfo{pages}{e1003972} (\bibinfo{year}{2013}).

\bibitem[{\citenamefont{K{\"{a}}renlampi}(2014)}]{Karenlampi2014}
\bibinfo{author}{\bibfnamefont{P.~P.} \bibnamefont{K{\"{a}}renlampi}},
  \bibinfo{journal}{Eur. Phys. J. E} \textbf{\bibinfo{volume}{37}},
  \bibinfo{pages}{56} (\bibinfo{year}{2014}).

\bibitem[{\citenamefont{Mathiesen et~al.}(2011)\citenamefont{Mathiesen,
  Mitarai, Sneppen, and Trusina}}]{Mathiesen2011}
\bibinfo{author}{\bibfnamefont{J.}~\bibnamefont{Mathiesen}},
  \bibinfo{author}{\bibfnamefont{N.}~\bibnamefont{Mitarai}},
  \bibinfo{author}{\bibfnamefont{K.}~\bibnamefont{Sneppen}}, \bibnamefont{and}
  \bibinfo{author}{\bibfnamefont{A.}~\bibnamefont{Trusina}},
  \bibinfo{journal}{Phys. Rev. Lett.} \textbf{\bibinfo{volume}{107}},
  \bibinfo{pages}{188101} (\bibinfo{year}{2011}).

\bibitem[{\citenamefont{Bagrow and Brockmann}(2013)}]{Bagrow2013}
\bibinfo{author}{\bibfnamefont{J.~P.} \bibnamefont{Bagrow}} \bibnamefont{and}
  \bibinfo{author}{\bibfnamefont{D.}~\bibnamefont{Brockmann}},
  \bibinfo{journal}{Phys. Rev. X} \textbf{\bibinfo{volume}{3}},
  \bibinfo{pages}{021016} (\bibinfo{year}{2013}).

\bibitem[{\citenamefont{Smith}(2007)}]{Smith2007}
\bibinfo{author}{\bibfnamefont{V.~H.} \bibnamefont{Smith}},
  \bibinfo{journal}{FEMS Microbiol. Ecol.} \textbf{\bibinfo{volume}{62}},
  \bibinfo{pages}{181} (\bibinfo{year}{2007}).

\bibitem[{\citenamefont{Vallina et~al.}(2014)\citenamefont{Vallina, Follows,
  Dutkiewicz, Montoya, Cermeno, and Loreau}}]{Vallina2014}
\bibinfo{author}{\bibfnamefont{S.~M.} \bibnamefont{Vallina}},
  \bibinfo{author}{\bibfnamefont{M.~J.} \bibnamefont{Follows}},
  \bibinfo{author}{\bibfnamefont{S.}~\bibnamefont{Dutkiewicz}},
  \bibinfo{author}{\bibfnamefont{J.~M.} \bibnamefont{Montoya}},
  \bibinfo{author}{\bibfnamefont{P.}~\bibnamefont{Cermeno}}, \bibnamefont{and}
  \bibinfo{author}{\bibfnamefont{M.}~\bibnamefont{Loreau}},
  \bibinfo{journal}{Nat. Commun.} \textbf{\bibinfo{volume}{5}},
  \bibinfo{pages}{4299} (\bibinfo{year}{2014}).

\bibitem[{\citenamefont{Nathan et~al.}(2016)\citenamefont{Nathan, Osem,
  Shachak, and Meron}}]{Nathan2015}
\bibinfo{author}{\bibfnamefont{J.}~\bibnamefont{Nathan}},
  \bibinfo{author}{\bibfnamefont{Y.}~\bibnamefont{Osem}},
  \bibinfo{author}{\bibfnamefont{M.}~\bibnamefont{Shachak}}, \bibnamefont{and}
  \bibinfo{author}{\bibfnamefont{E.}~\bibnamefont{Meron}}, \bibinfo{journal}{J.
  Ecol.} \textbf{\bibinfo{volume}{104}}, \bibinfo{pages}{419}
  (\bibinfo{year}{2016}).

\bibitem[{SM()}]{SM}
\emph{\bibinfo{title}{See supplemental material (sm) at [url will be inserted
  by publisher] for more information, definitions and results.}}

\bibitem[{\citenamefont{Thompson}(1998)}]{Thompson1998}
\bibinfo{author}{\bibfnamefont{J.~N.} \bibnamefont{Thompson}},
  \bibinfo{journal}{Trends Ecol. Evol.} \textbf{\bibinfo{volume}{13}},
  \bibinfo{pages}{329} (\bibinfo{year}{1998}).

\bibitem[{\citenamefont{Thorpe et~al.}(2011)\citenamefont{Thorpe, Aschehoug,
  Atwater, and Callaway}}]{Thorpe2011}
\bibinfo{author}{\bibfnamefont{A.~S.} \bibnamefont{Thorpe}},
  \bibinfo{author}{\bibfnamefont{E.~T.} \bibnamefont{Aschehoug}},
  \bibinfo{author}{\bibfnamefont{D.~Z.} \bibnamefont{Atwater}},
  \bibnamefont{and} \bibinfo{author}{\bibfnamefont{R.~M.}
  \bibnamefont{Callaway}}, \bibinfo{journal}{J. Ecol.}
  \textbf{\bibinfo{volume}{99}}, \bibinfo{pages}{729} (\bibinfo{year}{2011}).

\bibitem[{\citenamefont{Bergstrom and Kerr}(2015)}]{Bergstrom2015}
\bibinfo{author}{\bibfnamefont{C.~T.} \bibnamefont{Bergstrom}}
  \bibnamefont{and} \bibinfo{author}{\bibfnamefont{B.}~\bibnamefont{Kerr}},
  \bibinfo{journal}{Nature} \textbf{\bibinfo{volume}{521}},
  \bibinfo{pages}{431} (\bibinfo{year}{2015}).

\bibitem[{\citenamefont{Thompson}(1999)}]{Thompson1999}
\bibinfo{author}{\bibfnamefont{J.~N.} \bibnamefont{Thompson}},
  \bibinfo{journal}{Science (80-. ).} \textbf{\bibinfo{volume}{284}},
  \bibinfo{pages}{2116} (\bibinfo{year}{1999}).

\bibitem[{\citenamefont{Hubbell}(2001)}]{Hubbell2001}
\bibinfo{author}{\bibfnamefont{S.~P.} \bibnamefont{Hubbell}},
  \emph{\bibinfo{title}{The Unified Neutral Theory of Biodiversity and
  Biogeography}} (\bibinfo{publisher}{Princeton University Press},
  \bibinfo{address}{Princeton}, \bibinfo{year}{2001}).

\bibitem[{\citenamefont{Muller}(1932)}]{Muller1932}
\bibinfo{author}{\bibfnamefont{H.~J.} \bibnamefont{Muller}},
  \bibinfo{journal}{Am. Nat.} \textbf{\bibinfo{volume}{66}},
  \bibinfo{pages}{118} (\bibinfo{year}{1932}).

\bibitem[{\citenamefont{Maddamsetti et~al.}(2015)\citenamefont{Maddamsetti,
  Lenski, and Barrick}}]{Maddamsetti2015}
\bibinfo{author}{\bibfnamefont{R.}~\bibnamefont{Maddamsetti}},
  \bibinfo{author}{\bibfnamefont{R.~E.} \bibnamefont{Lenski}},
  \bibnamefont{and} \bibinfo{author}{\bibfnamefont{J.~E.}
  \bibnamefont{Barrick}}, \bibinfo{journal}{Genetics}
  \textbf{\bibinfo{volume}{200}}, \bibinfo{pages}{619} (\bibinfo{year}{2015}).

\bibitem[{\citenamefont{Szolnoki et~al.}(2014)\citenamefont{Szolnoki, Mobilia,
  Jiang, Szczesny, Rucklidge, and Perc}}]{Szolnoki2014}
\bibinfo{author}{\bibfnamefont{A.}~\bibnamefont{Szolnoki}},
  \bibinfo{author}{\bibfnamefont{M.}~\bibnamefont{Mobilia}},
  \bibinfo{author}{\bibfnamefont{L.-L.} \bibnamefont{Jiang}},
  \bibinfo{author}{\bibfnamefont{B.}~\bibnamefont{Szczesny}},
  \bibinfo{author}{\bibfnamefont{A.~M.} \bibnamefont{Rucklidge}},
  \bibnamefont{and} \bibinfo{author}{\bibfnamefont{M.}~\bibnamefont{Perc}},
  \bibinfo{journal}{J. R. Soc. Interface} \textbf{\bibinfo{volume}{11}},
  \bibinfo{pages}{20140735} (\bibinfo{year}{2014}).

\bibitem[{\citenamefont{Sinervo and Lively}(1996)}]{Sinervo1996}
\bibinfo{author}{\bibfnamefont{B.}~\bibnamefont{Sinervo}} \bibnamefont{and}
  \bibinfo{author}{\bibfnamefont{C.~M.} \bibnamefont{Lively}},
  \bibinfo{journal}{Nature} \textbf{\bibinfo{volume}{380}},
  \bibinfo{pages}{240} (\bibinfo{year}{1996}).

\bibitem[{\citenamefont{Lankau and Strauss}(2007)}]{Lankau2007}
\bibinfo{author}{\bibfnamefont{R.~a.} \bibnamefont{Lankau}} \bibnamefont{and}
  \bibinfo{author}{\bibfnamefont{S.~Y.} \bibnamefont{Strauss}},
  \bibinfo{journal}{Science (80-. ).} \textbf{\bibinfo{volume}{317}},
  \bibinfo{pages}{1561} (\bibinfo{year}{2007}).

\bibitem[{\citenamefont{Mitarai et~al.}(2012)\citenamefont{Mitarai, Mathiesen,
  and Sneppen}}]{Mitarai2012}
\bibinfo{author}{\bibfnamefont{N.}~\bibnamefont{Mitarai}},
  \bibinfo{author}{\bibfnamefont{J.}~\bibnamefont{Mathiesen}},
  \bibnamefont{and} \bibinfo{author}{\bibfnamefont{K.}~\bibnamefont{Sneppen}},
  \bibinfo{journal}{Phys. Rev. E} \textbf{\bibinfo{volume}{86}},
  \bibinfo{pages}{011929} (\bibinfo{year}{2012}).

\bibitem[{\citenamefont{Shtilerman et~al.}(2015)\citenamefont{Shtilerman,
  Kessler, and Shnerb}}]{Shtilerman2015}
\bibinfo{author}{\bibfnamefont{E.}~\bibnamefont{Shtilerman}},
  \bibinfo{author}{\bibfnamefont{D.~A.} \bibnamefont{Kessler}},
  \bibnamefont{and} \bibinfo{author}{\bibfnamefont{N.~M.}
  \bibnamefont{Shnerb}}, \bibinfo{journal}{J. Theor. Biol.}
  \textbf{\bibinfo{volume}{383}}, \bibinfo{pages}{138} (\bibinfo{year}{2015}).

\bibitem[{\citenamefont{Vandewalle and Ausloos}(1995)}]{Vandewalle1995}
\bibinfo{author}{\bibfnamefont{N.}~\bibnamefont{Vandewalle}} \bibnamefont{and}
  \bibinfo{author}{\bibfnamefont{M.}~\bibnamefont{Ausloos}},
  \bibinfo{journal}{Europhys. Lett.} \textbf{\bibinfo{volume}{32}},
  \bibinfo{pages}{613} (\bibinfo{year}{1995}).

\bibitem[{\citenamefont{Huisman et~al.}(2001)\citenamefont{Huisman, Johansson,
  Folmer, and Weissing}}]{Huisman2001}
\bibinfo{author}{\bibfnamefont{J.}~\bibnamefont{Huisman}},
  \bibinfo{author}{\bibfnamefont{A.~M.} \bibnamefont{Johansson}},
  \bibinfo{author}{\bibfnamefont{E.~O.} \bibnamefont{Folmer}},
  \bibnamefont{and} \bibinfo{author}{\bibfnamefont{F.~J.}
  \bibnamefont{Weissing}}, \bibinfo{journal}{Ecol. Lett.}
  \textbf{\bibinfo{volume}{4}}, \bibinfo{pages}{408} (\bibinfo{year}{2001}).

\bibitem[{\citenamefont{Beardmore et~al.}(2011)\citenamefont{Beardmore, Gudelj,
  Lipson, and Hurst}}]{Beardmore2011}
\bibinfo{author}{\bibfnamefont{R.~E.} \bibnamefont{Beardmore}},
  \bibinfo{author}{\bibfnamefont{I.}~\bibnamefont{Gudelj}},
  \bibinfo{author}{\bibfnamefont{D.~A.} \bibnamefont{Lipson}},
  \bibnamefont{and} \bibinfo{author}{\bibfnamefont{L.~D.} \bibnamefont{Hurst}},
  \bibinfo{journal}{Nature} \textbf{\bibinfo{volume}{472}},
  \bibinfo{pages}{342} (\bibinfo{year}{2011}).

\end{thebibliography}

\end{document}